\documentclass[final,1p,twocolumn]{elsarticle}

\usepackage{hyperref}
\usepackage{amsmath}
\usepackage{makecell}
\usepackage{booktabs}
\usepackage{float}

\usepackage{adjustbox}
\usepackage{amssymb}
\usepackage{amsthm}

\usepackage{booktabs}
\usepackage{multirow}

\def\argmin{\operatornamewithlimits{arg\,min}}
\newcommand\norm[1]{\left\lVert#1\right\rVert}

\usepackage{color}

\journal{Neural Networks}

\begin{document}

\begin{frontmatter}



\title{Learning Deformable Registration of Medical Images with Anatomical Constraints}

\author[]{Lucas Mansilla}
\author[]{Diego H. Milone}
\author[]{Enzo Ferrante}

\address{Research institute for signals, systems and computational intelligence, sinc(i)\\
FICH-UNL, CONICET\\ Santa Fe, Argentina}

\author{}

\address{}

\begin{abstract}
Deformable image registration is a fundamental problem in the field of medical image analysis. During the last years, we have witnessed the advent of deep learning-based image registration methods which achieve state-of-the-art performance, and drastically reduce the required computational time. However, little work has been done regarding how can we encourage our models to produce not only accurate, but also anatomically plausible results, which is still an open question in the field. In this work, we argue that incorporating anatomical priors in the form of global constraints into the learning process of these models, will further improve their performance and boost the realism of the warped images after registration. We learn global non-linear representations of image anatomy using segmentation masks, and employ them to constraint the registration process. The proposed AC-RegNet architecture is evaluated in the context of chest X-ray image registration using three different datasets, where the high anatomical variability makes the task extremely challenging. Our experiments show that the proposed anatomically constrained registration model produces more realistic and accurate results than state-of-the-art methods, demonstrating the potential of this approach.\\

\end{abstract}



\begin{keyword} medical image registration, convolutional neural networks, x-ray image analysis
\end{keyword}

\end{frontmatter}


\section{Introduction}
Deformable image registration is one of the pillar problems in the field of medical image analysis. Disease diagnosis, therapy planning, surgical and radiotherapy procedures are a few examples where image registration plays a crucial role. In the medical context, the problem consists in aligning two or more images coming from different patients, modalities, moments or view points. Such alignment is achieved by means of a deformation field, that warps the so called \textit{source} image, aligning it with the corresponding \textit{target} image.
\textcolor{black}{
Traditionally, image registration has been formulated as an optimization problem, where the objective function represents a similarity measure, which indicates how well the source image matches the target. 
}

Recently, the latest advances in machine learning allow us to conceive image registration under an entirely different paradigm. In particular, deep convolutional neural networks (CNN) have proved to outperform all existent strategies in other fundamental tasks of computer vision, like image segmentation \citep{long2015fully} and classification \citep{krizhevsky2012imagenet}. During the last years, we have witnessed the advent of deep learning-based image registration methods \citep{Li2018,Yang2017,Rohe2017,Sokooti2017,DeVos2017,de2019deep,Balakrishnan2018,Dalca2018,Estienne2019}, which achieve state-of-the-art performance, and drastically reduce the required computational time. These works have made a fundamental contribution by setting novel architectures for CNN-based deformable image registration (following supervised, unsupervised and semi-supervised training approaches). However, little work has been done regarding how can we encourage our models to produce not only accurate, but also anatomically plausible results, which is still an open question in the image registration community.

Image registration is employed in several medical tasks, e.g. when segmenting organs at risk in radiotherapy planning \citep{oh2017deformable}, co-registering MRI brain images prior to neuro-morphometry analysis \citep{gaser2016structural} or fusing pre and post-operative images to perform minimally invasive surgeries \citep{ferrante2017slice}). In these applications, producing anatomically plausible results is of paramount importance to guarantee the correctness of such medical and analytical procedures. Most of the existing registration methods impose smoothness constraints to the deformation fields or incorporate pixel-level losses into the objective functions to encourage anatomical plausibility. However, smooth deformation fields do not guarantee realistic results: in some cases, like moving organs for example, we require sharper deformations in the organ boundaries in order to preserve the anatomy. Regarding current pixel-based metrics (like Dice coefficient or Cross Entropy, for example), they do not consider the complete global context and therefore do not necessarily correlate with higher anatomical plausibility. The challenge we address is how CNN-based image registration models can produce accurate and anatomically plausible results after registration, that is, realistic results. 
Objectively quantifying the degree of realism of a medical image and incorporating it into the learning process of deep learning-based registration models is an extremely complex task. Here we propose to learn such measure from anatomical segmentation masks.


In this work, we argue that incorporating priors in the form of global anatomical constraints \citep{Oktay2017} into the learning process of deep learning-based registration models, will further improve the accuracy of the results and boost the realism of the warped images after registration. 
\textcolor{black}{
We address the question about how we can incorporate anatomical priors into deep learning-based image registration methods in order to obtain more realistic results. In that sense, our contributions are four-fold: (i) we extend, for the first time, the concept of anatomically constrained neural networks \citep{Oktay2017} to the image registration problem, (ii) we perform a deeper study of the complementarity between global and local loss functions defined over segmentation masks, (iii) we introduce the novel \textcolor{black}{anatomically constrained registration network (AC-RegNet)} architecture and validate it in the challenging task of X-ray chest image registration, comparing its performance with state-of-the-art existing methods and (iv) we showcase several application scenarios for AC-RegNet in the context of X-ray chest image analysis including multi-atlas segmentation, automatic quality control and pathology classification}\textcolor{black}{, where anatomical plausibility is highly relevant, especially when performing pathology classification using anatomical segmentation masks. In this case, the masks are obtained through registration-based label propagation. Thus, the anatomical plausibility of the deformed segmentation masks is crucial for the classification task.} We evaluate the proposed method in the context of X-ray chest imaging using three different datasets, including an interesting study about the behaviour of the global anatomical constraints when compared with a local metric. We show that the proposed method encourages the registration models to warp images in the space of anatomically plausible solutions while, at the same time, increasing the accuracy of the results. 

\section{Related works}
Inspired by \cite{Horn1980} and \cite{Lucas1981}, the research communities of computer vision and medical imaging have made major efforts towards developing more accurate and efficient registration methods. Since then, deformable image registration has been modelled in multiple ways (see \cite{Sotiras2013} for a comprehensive description), most of them posing image registration as an optimization problem, which in its general form can be formulated as
\begin{equation}
\label{eq:ClassicalRegistrationEq}
    \mathcal{\hat{T}} = \argmin_{\mathcal{T}}  \mathcal{M}(I \circ \mathcal{T}, J) + \mathcal{R}(\mathcal{T}),
\end{equation}

\noindent where $I$ is the source (moving) image, $J$ is the target (fixed) image, $\mathcal{T}$ parameterizes a spatial transformation that maps each point of the image $I$ to $J$, $\mathcal{M}$ corresponds to the criterion of (dis)similarity that quantifies the quality of the alignment between the warped source image $I \circ \mathcal{T}$ and the target image $J$, and $\mathcal{R}$ corresponds to the regularization term that imposes geometric constraints on the solution. In deformable image registration, the spatial transformation $\mathcal{T}$ is characterized by a deformation field, which represents the pixel displacements. The optimal transformation $\mathcal{\hat{T}}$ aligning $I$ with $J$ is computed by solving this minimization problem. \textcolor{black}{Classical non-learning based methods, like the one implemented in the popular toolbox SimpleElastix \citep{simpleelastix}, which will be used as baseline in this work, compute the optimal transformation by iteratively exploring the space of potential transformation parameters. In this case, a B-spline transformation is used to parameterize the deformation field, whose parameters are adjusted by minimizing the differences between the images. The main advantage of these methods is that they do not require a training process. Therefore, they can easily adapt to unseen images, being robust enough to be used in different image modalities and organs. However, when dealing with deformable registration, these algorithms are computationally expensive due to the high dimensionality of the parameter space, making the image registration process highly time consuming.}

Existing CNN-based image registration methods are usually classified as supervised or unsupervised, depending on whether or not they use ground truth deformation fields to compute the loss function during training. Inspired by the original FlowNet for vector flow estimation \citep{dosovitskiy2015flownet}, supervised CNN-based image registration methods \citep{Yang2017,Rohe2017,Sokooti2017} posed image registration as a regression problem. Given a pair of source and target images, they aim at regressing a deformation field that matches the ground-truth. One of the advantages of these methods is its independence with respect to image modalities: given a training dataset with pairs of images and their corresponding ground-truth deformations, it learns to map images to deformation fields without using any kind of similarity measure to compare them. However, getting such good datasets is a difficult task and makes these approaches impractical.

On the contrary, unsupervised CNN-based medical image registration \citep{Li2018,DeVos2017,de2019deep,Balakrishnan2018,Dalca2018,christodoulidis2018,balakrishnan2019voxelmorph} does not require ground-truth deformation fields. Instead, these methods (and the original CNN-based unsupervised optical flow estimation method \citep{ren2017unsupervised}) solve the registration process by minimizing a loss function based on the (dis)similarity $\mathcal{M}$ between the deformed source image and the target. They use a differentiable warping module similar that used in spatial transformers \citep{Jaderberg2015}, to warp the source image during the forward-pass, and allow the gradients flow back during backpropagation. In such way, the model is trained to produce deformation fields that minimize the similarity-based loss function. At test time, a single forward pass will return the deformation field. In this work, we will follow this strategy to construct a baseline architecture (referred RegNet throughout this text) that will serve as baseline when evaluating the impact of the proposed anatomically constrained registration method.

Various approaches were envisioned in the literature to improve the accuracy and realism of the registration methods by incorporating prior information (about image modalities, anatomy and structure) into the registration process. Two of the most common strategies are knowledge-based transformations, where the information is encoded within the deformation model \citep{Wouters2006,Glocker2009b,delPalomar2008} and segmentation-aware strategies, which directly incorporate segmentation priors to the registration process. 
The main disadvantage of knowleadge-based transformations is that they are highly domain specific, especially in methods like \cite{delPalomar2008}, which employ biomechanical models of specific organs to predict real deformations.
In this work, we focus on the second alternative. Several non-deep learning based approaches \citep{shakeri2016prior,ferrante2017deformable,ferrante2018weakly} were proposed to take advantage of such segmentations in the context of discrete graph-based image registration \citep{Paragios2016}.
In \cite{shakeri2016prior}, probabilistic priors are incorporated to the registration process through a new term in the energy function of the proposed discrete formulation. In \cite{ferrante2017deformable,ferrante2018weakly} segmentation masks are used to perform weak supervision when learning to aggregate standard similarity measures for image registration. However, these methods are orders of magnitude much slower than deep learning based image registration models.
The first multi-modal CNN-based image registration method proposed in \cite{hu2018weakly}, incorporates segmentation masks into the loss function of a weakly supervised approach to guide the learning process. They use a pixel-level similarity measure defined on the segmentation masks, that makes it possible to register images independently of their modality.
A similar pixel-level measure based on the Dice coefficient was incorporated in the VoxelMorph \citep{balakrishnan2019voxelmorph} and U-ResNet \citep{Estienne2019} frameworks, and used in tandem with a standard intensity based loss. Still, local pixel-level loss functions do not consider the global context and might produce similar values for anatomically plausible and non-plausible segmentations. 
In this work, we build on top of these ideas by regularizing the learning process using a global and non-linear representation of the underlying anatomy. We show that this global term is complementary to existent pixel-level loss functions. Moreover, in the context of X-ray chest image registration, we improve the performance of existent registration methods by a significant margin while producing more realistic images after deformation.

\section{Learning deformable image registration with anatomical constraints}
\begin{figure*}[th!]
\begin{center}
    \includegraphics[width=0.95\linewidth]{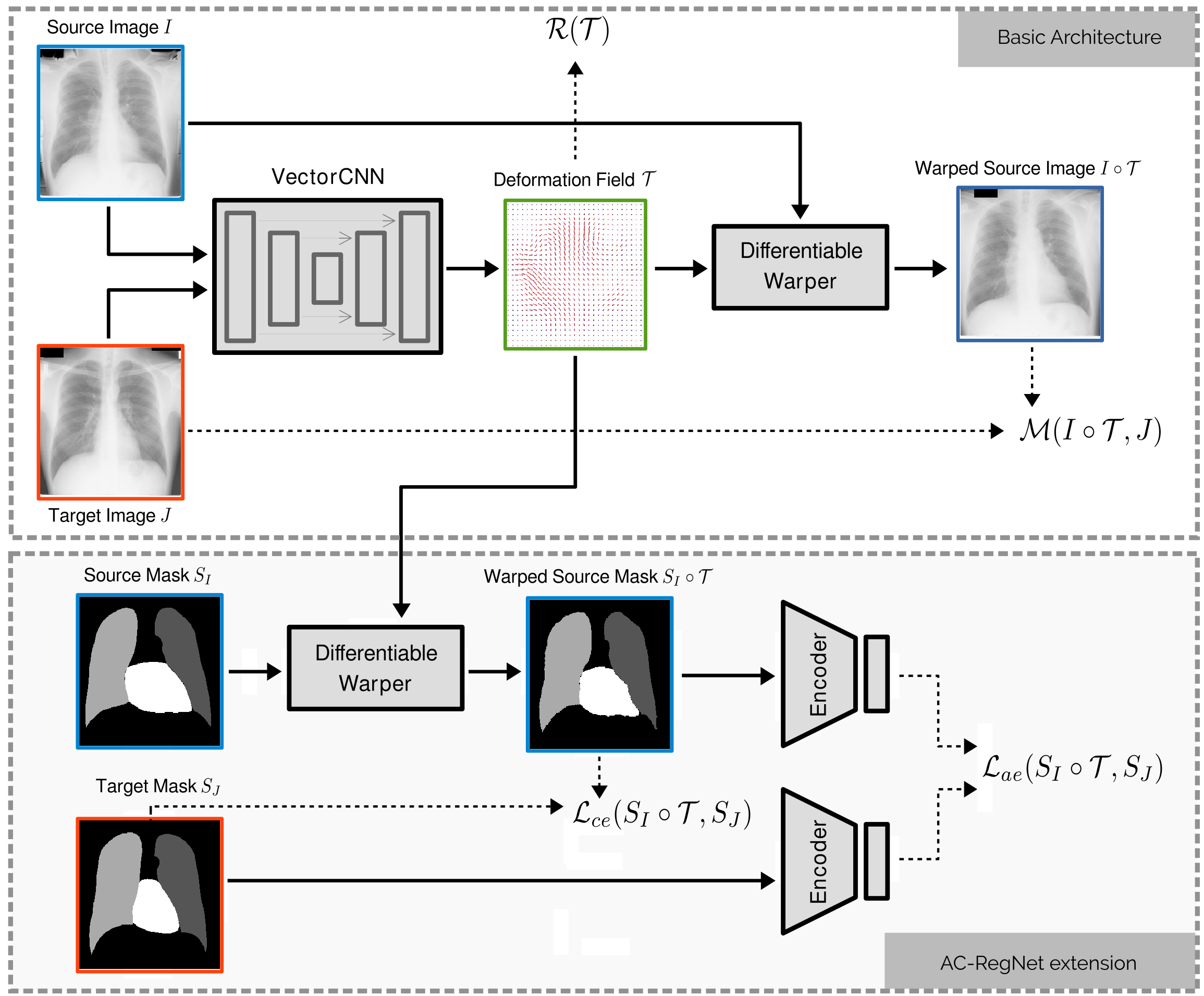}
\end{center}
    \caption{Architecture of the proposed AC-RegNet image registration model. \textcolor{black}{We combine a basic CNN architecture for image registration (upper box) with local ($\mathcal{L}_{ce}$) and a global ($\mathcal{L}_{ae}$) loss functions based on anatomical segmentations $S_I$, $S_J$ (lower box). We train the network VectorCNN to produce deformation fields $\mathcal{T}$ which minimize the differences between source ($I$) and target ($J$) images while, at the same time, ensuring anatomical correspondence after registration.}}
    \label{fig:Architecture}
\end{figure*}

\textcolor{black}{In this section, we provide a brief description of the existing basic CNN architecture used to perform unsupervised image registration, which constitutes the building blocks for the proposed model. }We then discuss how can we learn compact and non-linear representations of the image anatomy using denoising autoencoders (DAE) \citep{vincent2010stacked}, and how these representations can be introduced in the loss function to act as an anatomical regularizers, encouraging the learnt model to produce anatomically plausible images after deformation (see Figure \ref{fig:Architecture} for an overview of the proposed novel architecture).

\subsection{Basic architecture}
\label{sec:BasicArchitecture}
The basic CNN architecture for image registration is composed of two main modules. The first one (referred as VectorCNN in Figure \ref{fig:Architecture}) follows an encoder-decoder structure similar to that of U-Net \citep{Ronneberger2015}. Given a pair of source image $I$ and target image $J$ as input, VectorCNN predicts a deformation field $\mathcal{T} = \text{VectorCNN}(I, J; \Theta)$ where $\mathcal{T}: \mathcal{R}^n \rightarrow \mathcal{R}^n$ ($n$ beign the image dimensionality) and $\Theta$ corresponds to the network parameters learnt during training. Images $I$ and $J$ are concatenated and fed to the network as a single multi-channel image. VectorCNN processes the two images through a series of convolutional and pooling layers, and outputs a 2-channel filter map representing the 2D deformation field (3-channels in case we are dealing with 3D images). The second component is a differentiable warping module similar to that used in spatial transformer networks \citep{Jaderberg2015}, that uses $\mathcal{T}$ to deform the source image $I$, producing a warped image $I \circ \mathcal{T}$.

At the beginning of the training process, VectorCNN will produce random deformation fields $\mathcal{T}$. During training, the parameters $\Theta$ are adjusted so that the warped source image $I\circ \mathcal{T}$ minimizes the (dis)similarity criterion $\mathcal{M}$ with the target image $J$, in the same spirit that classic registration methods. In this work, we use the negative normalized cross correlation (NCC) to quantify image alignment. NCC has been previously used in the context of CNN-based registration \citep{Balakrishnan2018} and is a common choice when dealing with monomodal registration. We also consider a simple regularization term $\mathcal{R}(\mathcal{T})$ imposing smoothness to the deformation field by computing the total variation of the field \citep{Li2018,ferrante2018adaptability}. The basic loss function is therefore defined as
\begin{equation}
\label{eq:LossBasicArchitecture}
\mathcal{L}(I, J, \mathcal{T}) = \mathcal{M}(I \circ \mathcal{T}, J) + \lambda_{r} \mathcal{R}(\mathcal{T}),
\end{equation}

\noindent where $\lambda_{r}$ is a weighting factor for the total variation-based regularization term.

\subsection{Segmentation-aware local loss functions}
\label{sec:SegmentationLoss}
In order to augment the anatomical context provided to the network, we consider a simple initial strategy to include anatomical segmentations into the loss function by combining the aforementioned intensity-based loss $\mathcal{M}(I \circ \mathcal{T}, J)$, with a segmentation aware loss $\mathcal{L}_{ce}(S_I \circ \mathcal{T}, S_J)$. This new loss quantifies the alignment between a target anatomical segmentation mask $S_J$ and a warped version of a source segmentation mask $S_I$. The size of each segmentation mask is the same as that of the corresponding image, and the mask is formed by the elements (or pixels) $s_k \in \mathcal{C}$, where $\mathcal{C}$ is the set of classes. $\mathcal{L}_{ce}(S_I \circ \mathcal{T}, S_J)$ is implemented as the classical categorical cross-entropy defined at the pixel level on the one-hot encoded versions of $S_I$ and $S_J$. The segmentation-aware local loss function is thus defined as
\begin{eqnarray}
\label{eq:LossWithCE}
\mathcal{L}(I, J, S_I, S_J, \mathcal{T}) = \mathcal{M}(I \circ \mathcal{T}, J) + \lambda_{r} \mathcal{R}(\mathcal{T}) + \lambda_{ce} \mathcal{L}_{ce}(S_I \circ \mathcal{T}, S_J), 
\end{eqnarray}

\noindent where $\lambda_{ce}$ is a weighting factor for the additional term $\mathcal{L}_{ce}$. Note that segmentation masks $S_I, S_J$ are only required during training time to compute the loss function. At test time, a single pair of images will be fed into the network to produce a deformation field and no segmentation masks are required. 

\subsection{Auto-encoding global anatomical priors}
\label{sec:LearningPriors}
The local loss function $\mathcal{L}_{ce}(S_I \circ \mathcal{T}, S_J)$ defined in the previous section looks at pixel level predictions; therefore, it does not guarantee a good matching at the global scale between the deformed source and target anatomical masks. \textcolor{black}{The segmentation masks used in this work represent anatomical structures like lungs and heart. This is different from, for example, segmentation masks corresponding to pathological structures or lesions (like brain tumors or skin lesions), which are highly irregular both in terms of shape and topology. Even if anatomical structures present high variability among different patients, human organs maintain a high degree of regularity that we exploit to constraint the registration process.} We are interested in designing a loss function to analyze anatomical masks at a global scale, taking into account the anatomical plausibility of the deformed source mask when comparing it with the target mask. Since $\mathcal{L}_{ce}(S_I \circ \mathcal{T}, S_J)$ operates at the pixel level, the back-propagated gradients are parametrized only by pixel-wise individual probability terms and thus provide little global context \citep{Oktay2017}.

We learn a lower-dimensional representation of the anatomical segmentations using denoising autoencoders (DAE) \citep{vincent2010stacked}. Autoencoders are neural networks designed to learn a mapping from the input space $X$ to a novel, lower-dimensional representation $h$, that retains significant information about the input. These neural networks usually follow an encoder-decoder architecture
, where the encoding $h = \text{enc}(X)$ is extracted from an intermediate fully connected layer. This encoding contains significant information to decode the original input through a decoding phase $X \simeq \text{dec}(\text{enc}(X))$.

The model is trained to minimize the reconstruction error of the input masks, what results in maximizing a lower bound on the mutual information between the input $X$ and learnt representation $h$ \citep{vincent2010stacked}. In other words, the network is forced to store significant information (useful to reconstruct the original anatomical masks) into the learnt representation. A DAE considers noisy versions of the segmentation masks as input, and is trained to reconstruct clean versions of the corrupted input. This denoising effect, together with the bottleneck imposed by the encoder-decoder architecture, leads the model towards learning a manifold that captures the main variations in the data and maps similar segmentation masks into regions which are close in the manifold.

We train the DAE so that it minimizes the categorical cross-entropy defined over the one-hot encodings of our multi-organ anatomical masks. The noisy input segmentation masks for the DAE were constructed taking the clean segmentation masks and swapping the border pixels of the anatomical structures with the label of its left neighbor with a probability of 0.1. 
\textcolor{black}{Therefore, the learnt representation concentrates significant information about global anatomical properties of the organs, such as shape and topology, which can be used to introduce global anatomical priors in the learned registration model. In Section \ref{sec:Complementarity} we perform an experiment to better reflect the advantages of such representation when used as loss function, and discuss why it is beneficial to combine it with standard pixel-level losses.}


\subsection{AC-RegNet: Learning deformable image registration with anatomical constraints}
The novel \textcolor{black}{anatomically constrained registration network} (AC-RegNet) architecture is depicted in Figure \ref{fig:Architecture}. We combine the basic CNN architecture for image registration described in Section \ref{sec:BasicArchitecture} with the local segmentation-aware loss function $\mathcal{L}_{ce}$ (Section \ref{sec:SegmentationLoss}) and a new loss term based on the learnt anatomical representations (Section \ref{sec:LearningPriors}). This term will encourage global agreement between deformed source and target segmentation masks, ultimately resulting in more realistic and anatomically plausible images after warping.  The new term $\mathcal{L}_{ae}$ is defined as the squared Euclidean distance between the codes $h$ generated from the deformed source $S_I \circ \mathcal{T}$ and the corresponding target segmentation mask $S_J$ as:
\begin{eqnarray}
\label{eq:LossWithAE}
\mathcal{L}_{ae}(S_I \circ \mathcal{T}, S_J) =  \norm{enc(S_I \circ \mathcal{T}) - enc(S_J)}_2^2.
\end{eqnarray}

Note that both, the Euclidean norm and $\text{enc}(X)$ are differentiable operations and therefore $\mathcal{L}_{ae}$ is a differentiable loss. The final loss function for our AC-RegNet model considering both, local and global constraints, is given by:
\begin{eqnarray}
\label{eq:LossAC}
\mathcal{L}(I, J, S_I, S_J, \mathcal{T}) &=& \mathcal{M}(I \circ \mathcal{T}, J) + \lambda_{r} \mathcal{R}(\mathcal{T})  + \lambda_{ce} \mathcal{L}_{ce}(S_I \circ \mathcal{T}, S_J) + \\ && \lambda_{ae}  \mathcal{L}_{ae}(S_I \circ \mathcal{T}, S_J). \nonumber
\end{eqnarray}

The influence of the new term $\mathcal{L}_{ae}$ in the loss function is controlled with a weighting factor $\lambda_{ae}$. The main difference between $\mathcal{L}_{ae}$ and the pixel-level $\mathcal{L}_{ce}$ is that the first one acts at a global scale, better reflecting agreement in terms of anatomical shape variations. A deeper study about the complementarity of both losses is provided in Section \ref{sec:Complementarity}.


\subsubsection{Training the AC-RegNet model}
The training is organized in two stages. First, we train the autoencoder to learn a global and lower-dimensional representation of the anatomical structures using the segmentation masks. Second, we train the AC-RegNet model, by learning the parameters $\Theta$ of the VectorCNN that will produce the deformation field $\mathcal{T}$, considering the loss function defined in (\ref{eq:LossAC}). In this second stage, the parameters of the encoder model used to produce the codes $h = \text{enc}(S)$ are fixed. We highlight the fact that segmentation masks are used during training \textcolor{black}{(of both, the autoencoder and AC-RegNet models)} but, at test time, we only require the pair of images to be registered. \textcolor{black}{This is possible since the CNN which produces the deformation field given the pair of images (referred as "Basic Architecture" in Figure \ref{fig:Architecture}) does not use the segmentation masks. Instead, segmentation masks representing anatomical structures of interest in the corresponding images are used during training to enforce anatomical correspondences between the target and deformed source through the loss function defined in (\ref{eq:LossAC}). Once the AC-RegNet model is trained, the registration process of a given pair of images is performed by calculating the deformation field with VectorCNN and deforming the source image with the Warper module (see upper block in Figure \ref{fig:Architecture}), without the need of segmentations. The anatomical constraints are therefore introduced in the model during training.}

\section{Data and Experimental Setup}
\subsection{Image dataset}
The proposed registration model is evaluated in the context of inter-subject 2D chest X-ray image registration. Performing such task for different patients is challenging, since the anatomical variability between two different subjects can be really high. In our experiments, we use three image databases: the Japanese Society of Radiological Technology (JSRT) database \citep{JSRT}, the Montgomery County X-ray database and the Shenzhen Hospital X-ray database \citep{candemir2013lung,jaeger2013automatic}. These last two databases were created by the U.S. National Library of Medicine in collaboration with the Department of Health and Human Services, Montgomery County, Maryland, USA and the Shenzhen No.3 People’s Hospital, Guangdong Medical College, Shenzhen, China, respectively.

JSRT is a public database containing 247 PA chest X-ray images with and without lung nodules \textcolor{black}{(154 nodule and 93 non-nodule images)} of 2048x2048 pixels and a spacing of 0.175 mm/pixel. The Montgomery set contains 138 PA X-ray images with and without manifestations of tuberculosis \textcolor{black}{(80 normal and 58 pathological images)} of 4020x4892 or 4892x4020 pixels and a spacing of 0.0875 mm/pixel. The Shenzhen set contains 662 X-ray images with and without manifestations of tuberculosis \textcolor{black}{(326 normal and 336 pathological images)} in different sizes. Spacing is not provided, so we report results in pixel space when computing distance based measures like Hausdorff distance, \textcolor{black}{which measures the maximum distance between segmentation contours\footnote{We used the Hausdorff distance implementation available in MedPy package.}}. JSRT provides manual lung and heart segmentations for each image. Manual lung segmentations are available for Montgomery and Shenzhen sets. These segmentation masks will be used to learn the lower-dimensional representations and introduce anatomical context to the registration problem. 

The images and segmentations of the Montgomery and Shenzhen sets were preprocessed in order to obtain square images in the same spatial resolution. In each dataset, an image was taken as a reference image and resized by filling its shortest side with background color to make it square. Then, all the images of each dataset were registered against this image, taken as a reference image, through a similarity transform using SimpleElastix \citep{simpleelastix}\footnote{The configuration files used to run Elastix can be found online at \url{https://github.com/lucasmansilla/ACRN_Chest_X-ray_IA/tree/master/acregnet/config/JSRT/elastix}}, finally obtaining images of 4892x4892 pixels in the Montgomery set and 3000x3000 pixels in the Shenzhen set.

\label{sec:Experiments}
\subsection{Experimental setting}
We \textcolor{black}{randomly} divided the images of each dataset in 60\% training, 20\% validation and 20\% test. In the training stage, we sample random pairs of images from the training fold and built mini-batches of size 32. For testing, we sample $2 \times N$ random pairs of images from the test fold, where $N$ represents the number of images in that fold of the dataset.

In order to evaluate the performance of image registration algorithms we employ three metrics commonly used in the literature, which quantify the agreement between the warped source segmentation after registration and the target masks: (i) Dice Similarity Coefficient (DSC), which measures the overlapping between the segmentations \citep{dice}, (ii) Hausdorff Distance (HD), maximum distance between segmentation contours, and (iii) Average Symmetric Surface Distance (ASSD), computed as the average distance between the segmentation contours. DSC varies between 0 and 1, with 1 indicating a total correspondence between segmentations. HD and ASSD measure the distance between contours in milimeters, and lower values indicate better performance.

\subsection{Implementation details}
The proposed models were implemented in TensorFlow and trained with Adam optimizer considering learning rate of $10^{-3}$ and default TensorFlow values for the remaining optimization meta-parameters\footnote{Our code is available at \url{https://github.com/lucasmansilla/ACRN_Chest_X-ray_IA}. \textcolor{black}{Requirements and instructions are listed at \url{https://github.com/lucasmansilla/ACRN_Chest_X-ray_IA/tree/master/CLI_application/acregnet}}}. Models were trained until convergence and the weighting factors for the loss functions were chosen trough grid search using the validation fold, resulting in $\lambda_{r}=5\times10^{-5}$, $\lambda_{ce}=1$ and $\lambda_{ae}=10^{-1}$. A detailed description of the CNN architectures is provided in  \ref{sec:appendix}. 

\section{Results and Discussion}
\begin{figure*}[t!]
\begin{center}
   \includegraphics[width=1\linewidth]{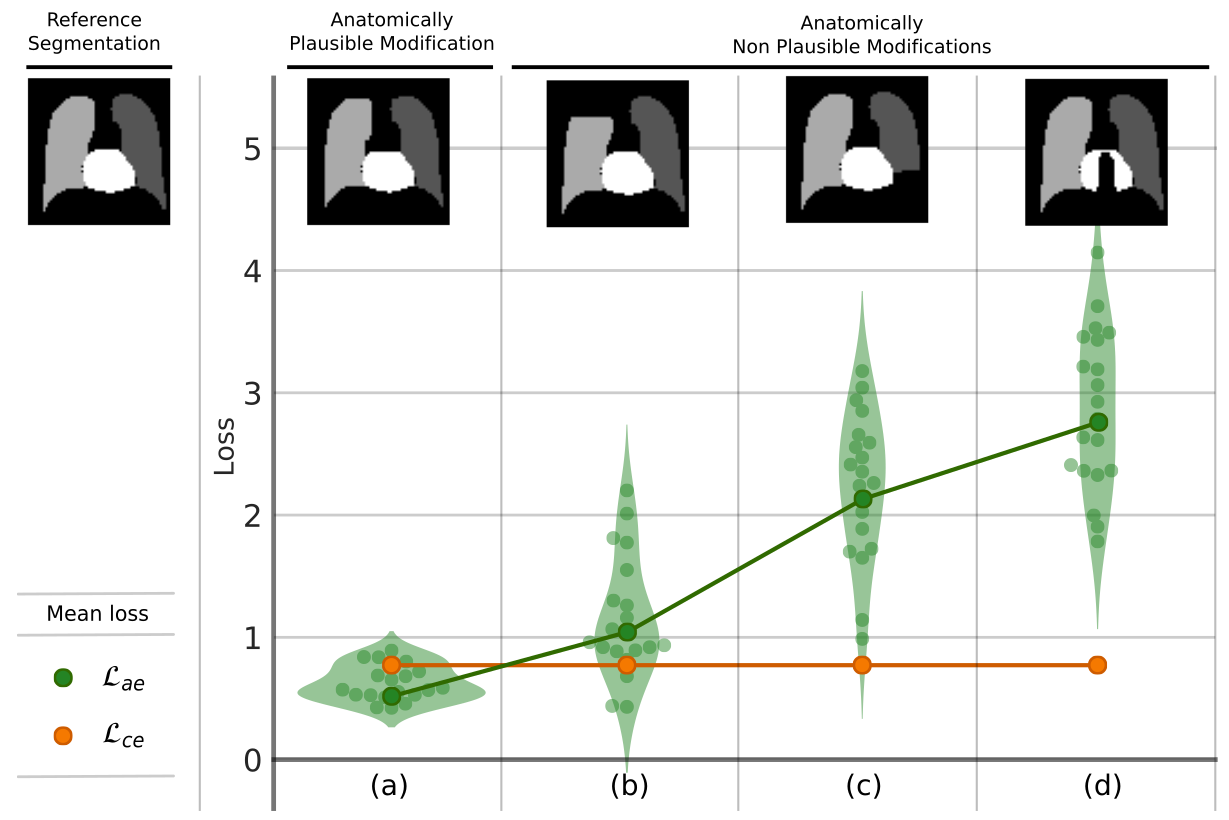}
\end{center}
    \caption{Comparison between the local loss $\mathcal{L}_{ce}$ defined at the pixel level, and the global loss $\mathcal{L}_{ae}$ based on the learnt representation, when comparing 20 random reference segmentations from JSRT dataset with its modified versions. The modified masks were obtained by manually setting 120 foreground pixels to background forming anatomically plausible (a) and non-plausible (b,c,d) versions. \textcolor{black}{The figure shows that $\mathcal{L}_{ae}$ encodes complementary information with respect to $\mathcal{L}_{ce}$}. Note that while the local $\mathcal{L}_{ce}$ remains constant, the global $\mathcal{L}_{ae}$ is much lower for the anatomically plausible cases. \textcolor{black}{This indicates that the proposed $\mathcal{L}_{ae}$ loss can discriminate different degrees of anatomical plausibility in cases where the standard pixel level functions indicate equivalent accuracy.}}
\label{fig:LossComparison}   
\end{figure*}
\subsection{Understanding the anatomical constraints}
\label{sec:Complementarity}
We perform a first experiment to compare the behaviour of the proposed global loss function ($\mathcal{L}_{ae}$) with the standard pixel-level loss ($\mathcal{L}_{ce}$), when comparing anatomically plausible and non-plausible segmentation masks. We take 20 random segmentation masks from our dataset, and generate 4 modified versions of each one by changing a constant number of pixels (120 pixels in our example) from the original segmentation (see images (a), (b), (c) and (d) in Figure \ref{fig:LossComparison}). While the segmentation mask (a) corresponds to an anatomically plausible version of the original mask (we just erode the mask by changing to background 120 pixels in the lungs and heart border), the other versions correspond to anatomically non-plausible masks (we remove blocks of 120 pixels representing complete parts of the lungs or heart). Remind that, in all these cases, a fixed number of pixels was changed. We then compute both losses $\mathcal{L}_{ae}$, $\mathcal{L}_{ce}$ and compare the reference segmentation with its modified versions. As expected, the local $\mathcal{L}_{ce}$ remained constant for the 4 cases, regardless of the place where the pixels are modified, since the number of non-agreeing pixels was 120 for all of them. However, when observing the behaviour of the global loss $\mathcal{L}_{ae}$, it returned a much lower value for the anatomically plausible case than for the non-plausible cases. Figure \ref{fig:LossComparison} shows the loss value for the 20 modified random masks following the same tendency: while the local $\mathcal{L}_{ce}$ remained constant in all cases, the global $\mathcal{L}_{ae}$ returned higher values for the non-plausible masks.

\begin{table*}[!t]
\begin{center}
\resizebox{\columnwidth}{!}{
\begin{tabular}{l l c c c}
\toprule
& & \multicolumn{3}{c}{Metric} \\
\cmidrule(l){3-5}
Dataset & Method & DSC & HD & ASSD \\
\midrule
& \textbf{AC-RegNet} & \textbf{0.943 (0.020)} & \textbf{17.973 (7.356)} & \textbf{3.340 (1.210)} \\
& AE-RegNet	& 0.934 (0.021)	& 19.464 (8.277) & 3.846 (1.320) \\
JSRT & CE-RegNet & 0.925 (0.025) & 21.973 (8.966) & 4.466 (1.553) \\
& RegNet & 0.809 (0.085) & 42.177 (19.751) & 11.229 (5.035) \\
& SimpleElastix & 0.846 (0.087) & 35.713 (18.180) & 9.028 (5.050) \\
\midrule
& \textbf{AC-RegNet} & \textbf{0.953 (0.017)} & \textbf{14.963 (7.910)} & \textbf{2.645 (0.957)} \\
& AE-RegNet	& 0.947 (0.019)	& 16.880 (8.621) & 2.981 (1.167) \\
Montgomery & CE-RegNet	& 0.929 (0.027)	& 33.425 (22.813) &	4.349 (1.945) \\
& RegNet	& 0.869 (0.052)	& 45.152 (35.702) & 8.078 (5.002) \\
& SimpleElastix	& 0.879 (0.073)	& 42.504 (27.480) &	7.136 (5.130) \\
\midrule
& \textbf{AC-RegNet} & \textbf{0.931 (0.027)} & \textbf{277.386 (182.207)} & \textbf{31.738 (15.891)} \\
& AE-RegNet	& 0.924 (0.032)	& 285.549 (179.823)	& 34.452 (18.259) \\
Shenzhen & CE-RegNet & 0.908 (0.039) & 325.958 (201.213) & 42.845 (23.560) \\
& RegNet & 0.830 (0.073) & 410.012 (225.783) & 73.758 (35.849) \\
& SimpleElastix & 0.883 (0.058) & 353.562 (217.423)	& 51.978 (30.299) \\
\bottomrule
\end{tabular}}
\caption{Mean and standard deviation of Dice Similarity Coefficient (DSC), Hausdorff Distance (HD) and Average Symmetric Surface Distance (ASSD) along all classes (left/right lung and heart) from JSRT, Montgomery and Shenzhen datasets. HD and ASSD for JSRT and Montgomery are expressed in milimeters, while Shenzen is expressed in pixels. Differences among the distributions for all pairs of method are statistically significant according to a paired Wilcoxon test considering Bonferroni correction. 
\textcolor{black}{Note that AC-RegNet achieves the best results, presenting the highest DSC and lowest HD and ASSD.}}
\label{tab:Metrics}
\end{center}
\end{table*}

This confirms our intuition about how $\mathcal{L}_{ae}$ encodes complementary information with respect to $\mathcal{L}_{ce}$. In the next section, we will see how this sensitivity to anatomical differences at the global scale can be exploited to improve the accuracy of our registration algorithm.

\begin{figure*}
\begin{center}
   \includegraphics[width=0.78\linewidth]{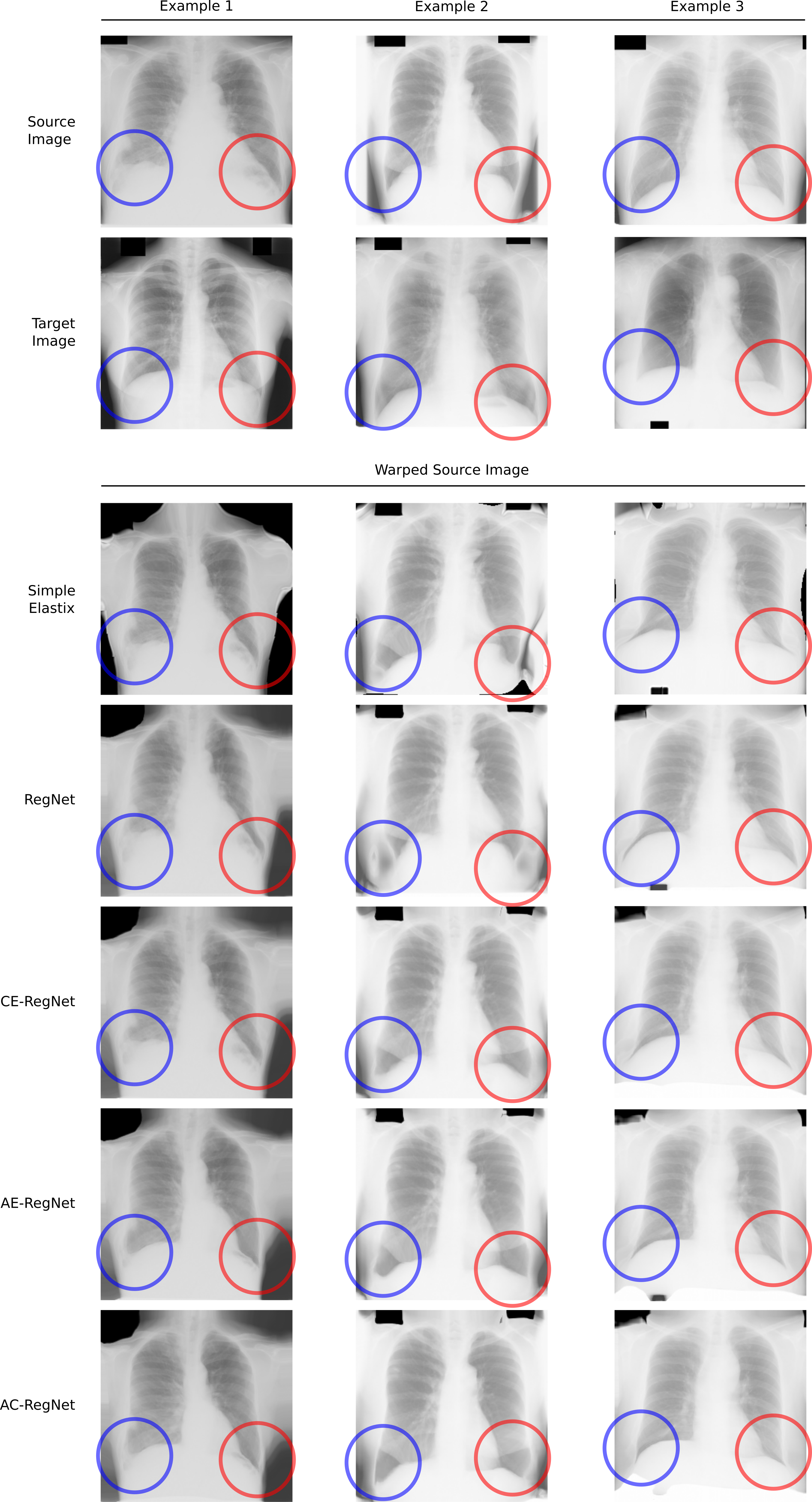}
\end{center}
    \caption{Visualization of the results after registering a pair of images. The blue and red circles highlight two of the areas of the lung anatomy that is better preserved by the AC-RegNet, when compared with the basic RegNet and the other segmentation-aware models. \textcolor{black}{Note, for example, that unlike AC-RegNet, other methods such as SimpleElastix (row 3) and RegNet (row 4), produce images where lungs are cut or stretched down.}}
\label{fig:qualitative}   
\end{figure*}

\subsection{Model comparison}
The proposed AC-RegNet model was compared with two initial baselines: SimpleElastix \citep{simpleelastix} and the baseline RegNet described in Section \ref{sec:BasicArchitecture}, which do not consider segmentation-aware loss functions during training. \textcolor{black}{SimpleElastix is a classic medical image registration toolbox, considered state-of-the-art and listed as one of the most popular software packages implementing iterative image registration during the last 20 years (see \cite{Vierger2016}). It has been recently used as baseline method not only for chest X-ray image registration \citep{ferrante2018adaptability} but also in a variety of applications (see for example \cite{de2019deep} or a complete list of publications in the Elastix website\footnote{List of publications using Elastix and their corresponding parameter files: \url{http://elastix.bigr.nl/wiki/index.php/Parameter_file_database}.}}).

We also include two segmentation-aware models, one considering only the local $\mathcal{L}_{ce}$ loss (referred as CE-RegNet \textcolor{black}{for Cross Entropy}) and another one considering only the global anatomical loss $\mathcal{L}_{ae}$ (referred as AE-RegNet \textcolor{black}{for Auto-Encoder}). The proposed model AC-RegNet considers a combination of both losses, as described in (\ref{eq:LossAC}). 
The quantitative results reported in Table \ref{tab:Metrics} show that our results are comparable to the state-of-the-art \citep{larrazabal2019anatomical,candemir2019review}. 
Statistical significance of the results between each pair of methods was verified by performing a Wilcoxon signed-rank test considering Bonferroni correction for each evaluation metric. In particular, when comparing the two most accurate methods (AC-Regnet and AE-RegNet) we obtained 
$p < 0.005$ 
in JSRT, $p < 0.05$ in Montgomery and $p < 0.005$ in Shenzhen dataset. Table \ref{tab:Metrics} shows that all segmentation-aware strategies (AC-RegNet, AE-RegNet and CE-RegNet) outperform baseline models (SimpleElastix and RegNet) by a significant margin, giving the highest DSC and lowest HD and ASSD values in each dataset. This already indicates that providing an anatomical context to the network helps to improve performance.
Moreover, using the combined local and global metrics (AC-RegNet) yields better performance than the individual cases. Figure \ref{fig:qualitative} illustrates the regularization effect produced by the AC-RegNet when compared with the other models. 
Note that registering images with AC-RegNet ensures a better anatomical correspondence between the organs, preventing unrealistic shapes in the lung areas as it happens with methods such as SimpleElastix and RegNet (where lungs are cut or stretched down).
These results confirm our previous study about the complementarity of both loss functions (see Section \ref{sec:Complementarity}), and the importance of considering global shape information on top of pixel-level descriptors to obtain more anatomically plausible results.

\subsection{Applications to X-ray image analysis}
In this section we aim at highlighting the potential of AC-RegNet in a variety of medical image analysis tasks. We show three different applications of the proposed method in X-ray images: (i) multi-atlas image segmentation, (ii) reverse classification accuracy (RCA) estimation \citep{valindria2017reverse} and (iii) representation learning for pathology classification. We use the well known NIH Chest-XRay14 dataset \citep{wang2017chestx} that includes 112,120 chest X-ray images labeled with 14 common thorax diseases according to an automatic natural language processing (NLP) analysis of the radiology reports. \\

\noindent \textbf{Multi-atlas image segmentation:} Anatomical segmentations are useful when performing disease classification and population analysis. The Chest-XRay14 is one of the largest medical datasets publicly available. However, it does not include anatomical segmentations. We used the AC-RegNet model to implement a multi-atlas segmentation model \citep{iglesias2015} and produce anatomical masks of lung and heart for all the images, which we are making publicly available\footnote{The resulting anatomical segmentation masks together with their corresponding RCA coefficient that estimates the quality of the segmentation can be downloaded from: \url{https://github.com/lucasmansilla/NIH_chest_xray14_segmentations}}. We follow a simple multi-atlas segmentation strategy \citep{mansilla2018}: given a target image, we take the 5 most similar images from the JSRT dataset (those which maximize the normalized cross correlation with that image) and apply AC-RegNet to register all of them to the target image space. We then transfer the JSRT segmentation labels by applying the resulting deformation field and fuse them using a simple majority voting mechanism. 

We believe that these segmentations are a valuable by-product contribution of our work, which may be used by the medical imaging community to perform further analysis based on the Chest-XRay14 dataset. We conducted automatic quality control to estimate the accuracy of the segmentation using RCA as described in the \textcolor{black}{following paragraph.}\\

\begin{figure*}[th!]
\begin{center}
    \includegraphics[width=1\linewidth]{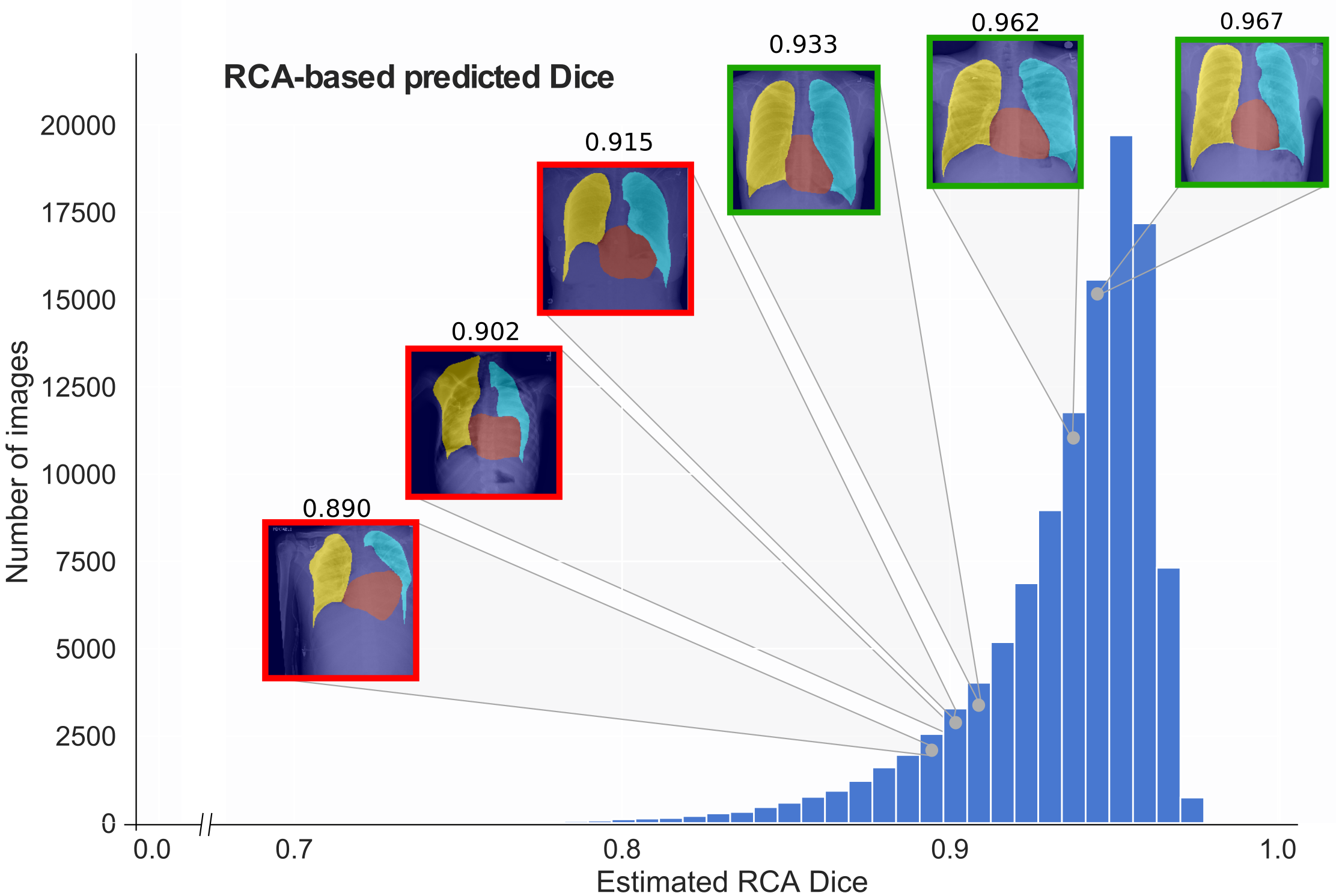}
\end{center}
    \caption{Histogram of estimated Dice coefficients for the resulting Chest-XRay14 segmentations. After visual inspection, we manually set a threshold of 0.92 to perform quality control and decide whether a segmentation meets (in green) or not (in red) the minimum quality standards. \textcolor{black}{We include some visual examples, which represent segmentation masks of various qualities obtained with the propose method, together with their corresponding RCA-based predicted Dice.}}
    \label{fig:hist}
\end{figure*}

\noindent \textbf{Reverse classification accuracy (RCA) estimation: } RCA is a framework for predicting the performance of a segmentation method on unseen data, first introduced in \cite{valindria2017reverse}. RCA takes the predicted segmentation from a new image to ``train" a reverse classifier (reverse in the sense that it is trained using a prediction), which is evaluated on a set of reference images with available ground truth. Such reverse classifier may take different forms, ranging from a random forest classifier to a single-atlas segmentation method based on image registration. The hypothesis is that if the prediction is correct, then the RCA classifier trained with that predicted segmentation will perform well in the reference images (those with ground truth). For a more detailed description of RCA see \cite{valindria2017reverse}.

Here we use AC-RegNet to implement RCA based on image registration, estimating the Dice coefficient of the Chest-XRay14 anatomical masks considering JSRT as reference images with ground-truth annotations. In such way, we provide an estimated quality index associated with every segmentation mask, that can be used to define if a given segmentation is to be trusted or not. After visual inspection, we set that threshold in 0.92. Figure \ref{fig:hist} shows the histogram of RCA  Dice coefficients together with some visual examples of segmentation masks below and above the minimum quality threshold.\\

\noindent \textbf{Cardiomegaly classification: } Cardiomegaly refers to an enlarged heart seen on any imaging test and can be diagnosed based on the cardiothoracic ratio (CR) \citep{dimopoulos2013cardiothoracic}. Since CR can be computed using the boundaries derived from heart and lung masks \citep{li2019automatic}, the anatomical segmentations should provide enough information to distinguish between healthy and pathological cases. We evaluated the discriminative power of the segmentations which meet the minimum quality requirements (RCA Dice $>$ 0.92) in the task of cardiomegaly vs healthy control classification. After quality control, we kept 87,870 from the original 112,120 images, out of which 2019 were labeled with cardiomegaly. We also sampled 2019 healthy patients with RCA Dice $>$ 0.92 to create a balanced dataset of 4,038 images including control and pathological.

 We perform 20-fold cross validation on the aforementioned dataset training a support vector machine (SVM) \citep{cortes1995support} with two alternative inputs based on the segmentation masks. As a first alternative, we applied principal component analysis (PCA) to reduce the dimensionality of a vectorized version of the segmentation masks\footnote{We employ the Scikit-learn ( \url{https://scikit-learn.org/} ) implementation of PCA and SVM in our experiments.}, keeping the 32 principal components and using them as features to train the SVM. Second, we employ the 32-dimensional representation learnt by the same autoencoder used to impose anatomical constraints to the AC-RegNet model. We trained the SVM using these two alternative representations and obtained accuracies of 0.77 and 0.79 respectively. \textcolor{black}{This
 suggests that the learnt representation encodes useful information that can be exploited in other medical imaging scenarios.}

\section{Conclusions}

In this paper, we introduced a new method to regularize CNN-based deformable image registration by considering global anatomical priors in the form of segmentation masks. Our method learns a non-linear and compact representation of the anatomy associated with medical images, and uses it to constraint the training process of standard CNN-based image registration architectures. 
Our method outperforms baseline approaches using 3 different datasets of X-ray images . What is more, we have used the to showcase different application scenarios where the proposed method can be used beyond standard image registration tasks, to perform multi-atlas segmentation, reverse classification accuracy estimation and pathology classification.

We provide a comprehensive evaluation of the AC-RegNet model in a challenging problem like chest X-ray image registration, including quantitative and qualitative results in three different datasets \textcolor{black}{(namely JSRT, Montgomery and Shenzhen). We also showcase three different application scenarios in the context of X-ray image analysis using the NIH Chest X-Ray dataset (composed of more than 100,000 images), where the proposed AC-RegNet was used to perform image segmentation, quality control and pathology detection. In conclusion, we found that the proposed global loss function encodes significant information about the anatomical plausibility of a deformed segmentation mask, which complements existent local losses defined at the pixel level. When used in tandem with intensity-based metrics and local losses defined on the segmentation masks, the global loss introduces additional constraints during training that encourage the registration model to produce more anatomically plausible images after deformation.}

The proposed model was applied in the context of 2D image registration, but extending it to 3D images is straightforward. In the future, we plan to validate our model in the context of brain 3D image registration, where anatomical structures can be clearly identified and used to constraint the training process. Moreover, as suggested in \cite{hu2018weakly}, CNN-based image registration methods considering segmentation masks can help to alleviate the challenging task of multi-modal registration. We plan to explore how AC-RegNet can be used to develop fast, reliable and realistic image registration methods for multi-modal scenarios.

\section{Acknowledgments}
EF is beneficiary of an AXA Research Fund grant. The authors gratefully acknowledge NVIDIA Corporation with the donation of the GPUs used for this research, and the support of UNL (CAID-PIC-50220140100084LI and 2016-082) and ANPCyT (PICT 2014-2627, 2018-03907, 2018-3384).
\appendix{}
\section{Detailed architectures}
\label{sec:appendix}

\begin{table*}[H]
\caption{Structure of the Autoencoder. BN: Batch Normalization. ReLU: Rectified Linear Unit. FC: Fully Connected. Up: Upsampling by a factor of 2 with nearest neighbor interpolation.}
\begin{center}
\resizebox{0.9\columnwidth}{!}{
\begin{tabular}{l c c c c c}
\toprule
\thead{\bf Layer} & 
\thead{\bf Number of \\ \bf Filters} & 
\thead{\bf Feature Maps \\ \bf Size \\ \bf (Height$\times$Width$\times$Channels)} & 
\thead{\bf Filter \\ Size\bf } & 
\thead{\bf Stride} &
\thead{\bf Padding} \\
\midrule
Input & & $64\times64\times4$ \\
\midrule
Conv $+$ BN & 16 & $32\times32\times16$ & $3\times3$ & $2\times2$ & $1\times1$ \\
ReLU & & $32\times32\times16$ \\
Conv $+$ BN & 16 & $32\times32\times16$ & $3\times3$ & $1\times1$ & $1\times1$ \\
ReLU & & $32\times32\times16$ \\
\midrule

Conv $+$ BN & 32 & $16\times16\times32$ & $3\times3$ & $2\times2$ & $1\times1$ \\
ReLU & & $16\times16\times32$ \\
Conv $+$ BN & 32 & $16\times16\times32$ & $3\times3$ & $1\times1$ & $1\times1$ \\
ReLU & & $16\times16\times32$ \\
\midrule

Conv $+$ BN & 1 & $8\times8\times1$ & $3\times3$ & $2\times2$ & $1\times1$ \\
ReLU & & $8\times8\times1$ \\
\midrule

FC & & $32$ \\
FC & & $64$ \\
ReLU & & $64$ \\
\midrule

Up $+$ Conv $+$ BN & 32 & $16\times16\times32$ & $3\times3$ & $1\times1$ & $1\times1$ \\
ReLU & & $16\times16\times32$ \\
Conv $+$ BN & 32 & $16\times16\times32$ & $3\times3$ & $1\times1$ & $1\times1$ \\
ReLU & & $16\times16\times32$ \\
\midrule

Up $+$ Conv $+$ BN & 16 & $32\times32\times16$ & $3\times3$ & $1\times1$ & $1\times1$ \\
ReLU & & $32\times32\times16$ \\
Conv $+$ BN & 16 & $32\times32\times16$ & $3\times3$ & $1\times1$ & $1\times1$ \\
ReLU & & $32\times32\times16$ \\
\midrule

Up $+$ Conv $+$ BN & 16 & $64\times64\times16$ & $3\times3$ & $1\times1$ & $1\times1$ \\
ReLU & & $64\times64\times16$ \\
Conv & 4 & $64\times64\times4$ & $3\times3$ & $1\times1$ & $1\times1$ \\
\bottomrule
\end{tabular}}
\end{center}
\label{tab:TableAutoencoder}
\end{table*}

\begin{table*}[H]
\caption{VectorCNN takes the source and target images as input (64x64), which are concatenated and feed into the network. It predicts a 2D deformation field, which has the same resolution as the input images.
References: BN: Batch Normalization. ELU: Exponential Linear Unit. Up: Upsampling by a factor of 2 with nearest neighbor interpolation.}
\begin{center}
\resizebox{0.7\columnwidth}{!}{
\begin{tabular}{l c c c c c}
\toprule
\thead{\bf Layer} & 
\thead{\bf Number of \\ \bf Filters} & 
\thead{\bf Feature Maps \\ \bf Size \\ \bf (Height$\times$Width$\times$Channels)} & 
\thead{\bf Filter \\ Size\bf } & 
\thead{\bf Stride} &
\thead{\bf Padding} \\
\midrule
Input & & $64\times64\times2$ \\
\midrule

Conv $+$ BN & 16 & $64\times64\times16$ & $3\times3$ & $1\times1$ & $1\times1$ \\
ELU & & $64\times64\times16$ \\
Conv $+$ BN (a) & 16 & $64\times64\times16$ & $3\times3$ & $1\times1$ & $1\times1$ \\
ELU & & $64\times64\times16$ \\
Average Pooling & & $32\times32\times16$ & $2\times2$ & $2\times2$ & $1\times1$ \\
\midrule

Conv $+$ BN & 32 & $32\times32\times32$ & $3\times3$ & $1\times1$ & $1\times1$ \\
ELU & & $32\times32\times32$ \\
Conv $+$ BN (b) & 32 & $32\times32\times32$ & $3\times3$ & $1\times1$ & $1\times1$ \\
ELU & & $32\times32\times32$ \\
Average Pooling & & $16\times16\times32$ & $2\times2$ & $2\times2$ & $1\times1$ \\
\midrule

Conv $+$ BN & 64 & $16\times16\times64$ & $3\times3$ & $1\times1$ & $1\times1$ \\
ELU & & $16\times16\times64$ \\
Conv $+$ BN (c) & 64 & $16\times16\times64$ & $3\times3$ & $1\times1$ & $1\times1$ \\
ELU & & $16\times16\times64$ \\
Average Pooling & & $8\times8\times64$ & $2\times2$ & $2\times2$ & $1\times1$ \\
\midrule

Conv $+$ BN & 128 & $8\times8\times128$ & $3\times3$ & $1\times1$ & $1\times1$ \\
ELU & & $8\times8\times128$ \\
Conv $+$ BN & 128 & $8\times8\times128$ & $3\times3$ & $1\times1$ & $1\times1$ \\
ELU & & $8\times8\times128$ \\
Dropout (50\%) & & $8\times8\times128$ \\
\midrule

Up $+$ Conv $+$ BN & 64 & $16\times16\times64$ & $3\times3$ & $1\times1$ & $1\times1$ \\
Concat with (c) & & $16\times16\times64$ \\
ELU & & $16\times16\times64$ \\
Conv $+$ BN & 64 & $16\times16\times64$ & $3\times3$ & $1\times1$ & $1\times1$ \\
ELU & & $16\times16\times64$ \\
Conv $+$ BN & 64 & $16\times16\times64$ & $3\times3$ & $1\times1$ & $1\times1$ \\
ELU & & $16\times16\times64$ \\
Dropout (50\%) & & $16\times16\times64$ \\
\midrule

Up $+$ Conv $+$ BN & 32 & $32\times32\times32$ & $3\times3$ & $1\times1$ & $1\times1$ \\
Concat with (b) & & $32\times32\times32$ \\
ELU & & $32\times32\times32$ \\
Conv $+$ BN & 32 & $32\times32\times32$ & $3\times3$ & $1\times1$ & $1\times1$ \\
ELU & & $32\times32\times32$ \\
Conv $+$ BN & 32 & $32\times32\times32$ & $3\times3$ & $1\times1$ & $1\times1$ \\
ELU & & $32\times32\times32$ \\
Dropout (50\%) & & $32\times32\times32$ \\
\midrule

Up $+$ Conv $+$ BN & 16 & $64\times64\times16$ & $3\times3$ & $1\times1$ & $1\times1$ \\
Concat with (a) & & $64\times64\times16$ \\
ELU & & $64\times64\times16$ \\
Conv $+$ BN & 16 & $64\times64\times16$ & $3\times3$ & $1\times1$ & $1\times1$ \\
ELU & & $64\times64\times16$ \\
Conv $+$ BN & 16 & $64\times64\times16$ & $3\times3$ & $1\times1$ & $1\times1$ \\
ELU & & $64\times64\times16$ \\
Conv $+$ BN & 2 & $64\times64\times2$ & $3\times3$ & $1\times1$ & $1\times1$ \\
\bottomrule
\end{tabular}}
\end{center}
\label{tab:TableVectorCNN}
\end{table*}

\bibliographystyle{elsarticle-harv} 
\bibliography{references}

\begin{thebibliography}{49}
\expandafter\ifx\csname natexlab\endcsname\relax\def\natexlab#1{#1}\fi
\providecommand{\url}[1]{\texttt{#1}}
\providecommand{\href}[2]{#2}
\providecommand{\path}[1]{#1}
\providecommand{\DOIprefix}{doi:}
\providecommand{\ArXivprefix}{arXiv:}
\providecommand{\URLprefix}{URL: }
\providecommand{\Pubmedprefix}{pmid:}
\providecommand{\doi}[1]{\href{http://dx.doi.org/#1}{\path{#1}}}
\providecommand{\Pubmed}[1]{\href{pmid:#1}{\path{#1}}}
\providecommand{\bibinfo}[2]{#2}
\ifx\xfnm\relax \def\xfnm[#1]{\unskip,\space#1}\fi
\bibitem[{Balakrishnan et~al.(2018)Balakrishnan, Zhao, Sabuncu, Guttag and
  Dalca}]{Balakrishnan2018}
\bibinfo{author}{Balakrishnan, G.}, \bibinfo{author}{Zhao, A.},
  \bibinfo{author}{Sabuncu, M.R.}, \bibinfo{author}{Guttag, J.},
  \bibinfo{author}{Dalca, A.V.}, \bibinfo{year}{2018}.
\newblock \bibinfo{title}{An unsupervised learning model for deformable medical
  image registration}, in: \bibinfo{booktitle}{IEEE CVPR Proceedings}, pp.
  \bibinfo{pages}{9252--9260}.
\bibitem[{Balakrishnan et~al.(2019)Balakrishnan, Zhao, Sabuncu, Guttag and
  Dalca}]{balakrishnan2019voxelmorph}
\bibinfo{author}{Balakrishnan, G.}, \bibinfo{author}{Zhao, A.},
  \bibinfo{author}{Sabuncu, M.R.}, \bibinfo{author}{Guttag, J.},
  \bibinfo{author}{Dalca, A.V.}, \bibinfo{year}{2019}.
\newblock \bibinfo{title}{Voxelmorph: a learning framework for deformable
  medical image registration}.
\newblock \bibinfo{journal}{IEEE TMI} .
\bibitem[{Candemir and Antani(2019)}]{candemir2019review}
\bibinfo{author}{Candemir, S.}, \bibinfo{author}{Antani, S.},
  \bibinfo{year}{2019}.
\newblock \bibinfo{title}{A review on lung boundary detection in chest x-rays}.
\newblock \bibinfo{journal}{IJCARS} \bibinfo{volume}{14},
  \bibinfo{pages}{563--576}.
\bibitem[{Candemir et~al.(2013)Candemir, Jaeger, Palaniappan, Musco, Singh,
  Xue, Karargyris, Antani, Thoma and McDonald}]{candemir2013lung}
\bibinfo{author}{Candemir, S.}, \bibinfo{author}{Jaeger, S.},
  \bibinfo{author}{Palaniappan, K.}, \bibinfo{author}{Musco, J.P.},
  \bibinfo{author}{Singh, R.K.}, \bibinfo{author}{Xue, Z.},
  \bibinfo{author}{Karargyris, A.}, \bibinfo{author}{Antani, S.},
  \bibinfo{author}{Thoma, G.}, \bibinfo{author}{McDonald, C.J.},
  \bibinfo{year}{2013}.
\newblock \bibinfo{title}{Lung segmentation in chest radiographs using
  anatomical atlases with nonrigid registration}.
\newblock \bibinfo{journal}{IEEE TMI} \bibinfo{volume}{33},
  \bibinfo{pages}{577--590}.
\bibitem[{Cortes and Vapnik(1995)}]{cortes1995support}
\bibinfo{author}{Cortes, C.}, \bibinfo{author}{Vapnik, V.},
  \bibinfo{year}{1995}.
\newblock \bibinfo{title}{Support-vector networks}.
\newblock \bibinfo{journal}{Machine learning} \bibinfo{volume}{20},
  \bibinfo{pages}{273--297}.
\bibitem[{Dalca et~al.(2018)Dalca, Balakrishnan, Guttag and
  Sabuncu}]{Dalca2018}
\bibinfo{author}{Dalca, A.V.}, \bibinfo{author}{Balakrishnan, G.},
  \bibinfo{author}{Guttag, J.}, \bibinfo{author}{Sabuncu, M.R.},
  \bibinfo{year}{2018}.
\newblock \bibinfo{title}{Unsupervised learning for fast probabilistic
  diffeomorphic registration}.
\newblock \bibinfo{journal}{arXiv preprint arXiv:1805.04605} .
\bibitem[{Del~Palomar et~al.(2008)Del~Palomar, Calvo, Herrero, L{\'o}pez and
  Doblar{\'e}}]{delPalomar2008}
\bibinfo{author}{Del~Palomar, A.P.}, \bibinfo{author}{Calvo, B.},
  \bibinfo{author}{Herrero, J.}, \bibinfo{author}{L{\'o}pez, J.},
  \bibinfo{author}{Doblar{\'e}, M.}, \bibinfo{year}{2008}.
\newblock \bibinfo{title}{A finite element model to accurately predict real
  deformations of the breast}.
\newblock \bibinfo{journal}{Medical Engineering \& Physics}
  \bibinfo{volume}{30}, \bibinfo{pages}{1089--1097}.
\bibitem[{Dice(1945)}]{dice}
\bibinfo{author}{Dice, L.R.}, \bibinfo{year}{1945}.
\newblock \bibinfo{title}{Measures of the amount of ecologic association
  between species}.
\newblock \bibinfo{journal}{Ecology} \bibinfo{volume}{26},
  \bibinfo{pages}{297--302}.
\bibitem[{Dimopoulos et~al.(2013)Dimopoulos, Giannakoulas, Bendayan, Liodakis,
  Petraco, Diller, Piepoli, Swan, Mullen, Best
  et~al.}]{dimopoulos2013cardiothoracic}
\bibinfo{author}{Dimopoulos, K.}, \bibinfo{author}{Giannakoulas, G.},
  \bibinfo{author}{Bendayan, I.}, \bibinfo{author}{Liodakis, E.},
  \bibinfo{author}{Petraco, R.}, \bibinfo{author}{Diller, G.P.},
  \bibinfo{author}{Piepoli, M.F.}, \bibinfo{author}{Swan, L.},
  \bibinfo{author}{Mullen, M.}, \bibinfo{author}{Best, N.}, et~al.,
  \bibinfo{year}{2013}.
\newblock \bibinfo{title}{Cardiothoracic ratio from postero-anterior chest
  radiographs: a simple, reproducible and independent marker of disease
  severity and outcome in adults with congenital heart disease}.
\newblock \bibinfo{journal}{International journal of cardiology}
  \bibinfo{volume}{166}, \bibinfo{pages}{453--457}.
\bibitem[{Dosovitskiy et~al.(2015)Dosovitskiy, Fischer, Ilg, Hausser, Hazirbas,
  Golkov, Van Der~Smagt, Cremers and Brox}]{dosovitskiy2015flownet}
\bibinfo{author}{Dosovitskiy, A.}, \bibinfo{author}{Fischer, P.},
  \bibinfo{author}{Ilg, E.}, \bibinfo{author}{Hausser, P.},
  \bibinfo{author}{Hazirbas, C.}, \bibinfo{author}{Golkov, V.},
  \bibinfo{author}{Van Der~Smagt, P.}, \bibinfo{author}{Cremers, D.},
  \bibinfo{author}{Brox, T.}, \bibinfo{year}{2015}.
\newblock \bibinfo{title}{Flownet: Learning optical flow with convolutional
  networks}, in: \bibinfo{booktitle}{IEEE ICCV Proceedings}, pp.
  \bibinfo{pages}{2758--2766}.
\bibitem[{Estienne et~al.(2019)Estienne, Vakalopoulou, Christodoulidis,
  Battistela, Lerousseau, Carre, Klausner, Sun, Robert, Mougiakakou
  et~al.}]{Estienne2019}
\bibinfo{author}{Estienne, T.}, \bibinfo{author}{Vakalopoulou, M.},
  \bibinfo{author}{Christodoulidis, S.}, \bibinfo{author}{Battistela, E.},
  \bibinfo{author}{Lerousseau, M.}, \bibinfo{author}{Carre, A.},
  \bibinfo{author}{Klausner, G.}, \bibinfo{author}{Sun, R.},
  \bibinfo{author}{Robert, C.}, \bibinfo{author}{Mougiakakou, S.}, et~al.,
  \bibinfo{year}{2019}.
\newblock \bibinfo{title}{U-resnet: Ultimate coupling of registration and
  segmentation with deep nets}, in: \bibinfo{booktitle}{MICCAI},
  \bibinfo{organization}{Springer}. pp. \bibinfo{pages}{310--319}.
\bibitem[{Ferrante et~al.(2017)Ferrante, Dokania, Marini and
  Paragios}]{ferrante2017deformable}
\bibinfo{author}{Ferrante, E.}, \bibinfo{author}{Dokania, P.K.},
  \bibinfo{author}{Marini, R.}, \bibinfo{author}{Paragios, N.},
  \bibinfo{year}{2017}.
\newblock \bibinfo{title}{Deformable registration through learning of
  context-specific metric aggregation}, in: \bibinfo{booktitle}{MLMI},
  \bibinfo{organization}{Springer}. pp. \bibinfo{pages}{256--265}.
\bibitem[{Ferrante et~al.(2018a)Ferrante, Dokania, Silva and
  Paragios}]{ferrante2018weakly}
\bibinfo{author}{Ferrante, E.}, \bibinfo{author}{Dokania, P.K.},
  \bibinfo{author}{Silva, R.M.}, \bibinfo{author}{Paragios, N.},
  \bibinfo{year}{2018}a.
\newblock \bibinfo{title}{Weakly-supervised learning of metric aggregations for
  deformable image registration}.
\newblock \bibinfo{journal}{J-BHI} .
\bibitem[{Ferrante et~al.(2018b)Ferrante, Oktay, Glocker and
  Milone}]{ferrante2018adaptability}
\bibinfo{author}{Ferrante, E.}, \bibinfo{author}{Oktay, O.},
  \bibinfo{author}{Glocker, B.}, \bibinfo{author}{Milone, D.H.},
  \bibinfo{year}{2018}b.
\newblock \bibinfo{title}{On the adaptability of unsupervised cnn-based
  deformable image registration to unseen image domains}, in:
  \bibinfo{booktitle}{MLMI}, \bibinfo{organization}{Springer}. pp.
  \bibinfo{pages}{294--302}.
\bibitem[{Ferrante and Paragios(2017)}]{ferrante2017slice}
\bibinfo{author}{Ferrante, E.}, \bibinfo{author}{Paragios, N.},
  \bibinfo{year}{2017}.
\newblock \bibinfo{title}{Slice-to-volume medical image registration: A
  survey}.
\newblock \bibinfo{journal}{Medical Image Analysis} \bibinfo{volume}{39},
  \bibinfo{pages}{101--123}.
\bibitem[{Gaser(2016)}]{gaser2016structural}
\bibinfo{author}{Gaser, C.}, \bibinfo{year}{2016}.
\newblock \bibinfo{title}{Structural mri: Morphometry}, in:
  \bibinfo{booktitle}{Neuroeconomics}. \bibinfo{publisher}{Springer}, pp.
  \bibinfo{pages}{399--409}.
\bibitem[{Glocker et~al.(2009)Glocker, Komodakis, Navab, Tziritas and
  Paragios}]{Glocker2009b}
\bibinfo{author}{Glocker, B.}, \bibinfo{author}{Komodakis, N.},
  \bibinfo{author}{Navab, N.}, \bibinfo{author}{Tziritas, G.},
  \bibinfo{author}{Paragios, N.}, \bibinfo{year}{2009}.
\newblock \bibinfo{title}{{Dense Registration with Deformation Priors}}, in:
  \bibinfo{booktitle}{IPMI}, pp. \bibinfo{pages}{540--551}.
\bibitem[{Horn and Schunck.(1980)}]{Horn1980}
\bibinfo{author}{Horn, B.K.}, \bibinfo{author}{Schunck., B.G.},
  \bibinfo{year}{1980}.
\newblock \bibinfo{title}{{Determining Optical Flow}}.
\newblock \bibinfo{journal}{Artificial Intelligence} \bibinfo{volume}{17},
  \bibinfo{pages}{185--203}.
\bibitem[{Hu et~al.(2018)Hu, Modat, Gibson, Li, Ghavami, Bonmati, Wang,
  Bandula, Moore, Emberton et~al.}]{hu2018weakly}
\bibinfo{author}{Hu, Y.}, \bibinfo{author}{Modat, M.}, \bibinfo{author}{Gibson,
  E.}, \bibinfo{author}{Li, W.}, \bibinfo{author}{Ghavami, N.},
  \bibinfo{author}{Bonmati, E.}, \bibinfo{author}{Wang, G.},
  \bibinfo{author}{Bandula, S.}, \bibinfo{author}{Moore, C.M.},
  \bibinfo{author}{Emberton, M.}, et~al., \bibinfo{year}{2018}.
\newblock \bibinfo{title}{Weakly-supervised convolutional neural networks for
  multimodal image registration}.
\newblock \bibinfo{journal}{Medical Image Analysis} \bibinfo{volume}{49},
  \bibinfo{pages}{1--13}.
\bibitem[{Iglesias and Sabuncu(2015)}]{iglesias2015}
\bibinfo{author}{Iglesias, J.E.}, \bibinfo{author}{Sabuncu, M.R.},
  \bibinfo{year}{2015}.
\newblock \bibinfo{title}{Multi-atlas segmentation of biomedical images: a
  survey}.
\newblock \bibinfo{journal}{Medical Image Analysis} \bibinfo{volume}{24},
  \bibinfo{pages}{205--219}.
\bibitem[{Jaderberg et~al.(2015)}]{Jaderberg2015}
\bibinfo{author}{Jaderberg, M.}, et~al., \bibinfo{year}{2015}.
\newblock \bibinfo{title}{Spatial transformer networks}, in:
  \bibinfo{booktitle}{NIPS}, pp. \bibinfo{pages}{2017--2025}.
\bibitem[{Jaeger et~al.(2013)Jaeger, Karargyris, Candemir, Folio, Siegelman,
  Callaghan, Xue, Palaniappan, Singh, Antani et~al.}]{jaeger2013automatic}
\bibinfo{author}{Jaeger, S.}, \bibinfo{author}{Karargyris, A.},
  \bibinfo{author}{Candemir, S.}, \bibinfo{author}{Folio, L.},
  \bibinfo{author}{Siegelman, J.}, \bibinfo{author}{Callaghan, F.},
  \bibinfo{author}{Xue, Z.}, \bibinfo{author}{Palaniappan, K.},
  \bibinfo{author}{Singh, R.K.}, \bibinfo{author}{Antani, S.}, et~al.,
  \bibinfo{year}{2013}.
\newblock \bibinfo{title}{Automatic tuberculosis screening using chest
  radiographs}.
\newblock \bibinfo{journal}{IEEE TMI} \bibinfo{volume}{33},
  \bibinfo{pages}{233--245}.
\bibitem[{Krizhevsky et~al.(2012)Krizhevsky, Sutskever and
  Hinton}]{krizhevsky2012imagenet}
\bibinfo{author}{Krizhevsky, A.}, \bibinfo{author}{Sutskever, I.},
  \bibinfo{author}{Hinton, G.E.}, \bibinfo{year}{2012}.
\newblock \bibinfo{title}{Imagenet classification with deep convolutional
  neural networks}, in: \bibinfo{booktitle}{NIPS}, pp.
  \bibinfo{pages}{1097--1105}.
\bibitem[{Larrazabal et~al.(2019)Larrazabal, Martinez and
  Ferrante}]{larrazabal2019anatomical}
\bibinfo{author}{Larrazabal, A.J.}, \bibinfo{author}{Martinez, C.},
  \bibinfo{author}{Ferrante, E.}, \bibinfo{year}{2019}.
\newblock \bibinfo{title}{Anatomical priors for image segmentation via
  post-processing with denoising autoencoders}.
\newblock \bibinfo{journal}{In Proceedings of MICCAI 2019} .
\bibitem[{Li and Fan(2018)}]{Li2018}
\bibinfo{author}{Li, H.}, \bibinfo{author}{Fan, Y.}, \bibinfo{year}{2018}.
\newblock \bibinfo{title}{Non-rigid image registration using self-supervised
  fully convolutional networks without training data}, in:
  \bibinfo{booktitle}{ISBI}, \bibinfo{organization}{IEEE}. pp.
  \bibinfo{pages}{1075--1078}.
\bibitem[{Li et~al.(2019)Li, Hou, Chen, Hao, An, Liang and
  Lu}]{li2019automatic}
\bibinfo{author}{Li, Z.}, \bibinfo{author}{Hou, Z.}, \bibinfo{author}{Chen,
  C.}, \bibinfo{author}{Hao, Z.}, \bibinfo{author}{An, Y.},
  \bibinfo{author}{Liang, S.}, \bibinfo{author}{Lu, B.}, \bibinfo{year}{2019}.
\newblock \bibinfo{title}{Automatic cardiothoracic ratio calculation with deep
  learning}.
\newblock \bibinfo{journal}{IEEE Access} \bibinfo{volume}{7},
  \bibinfo{pages}{37749--37756}.
\bibitem[{Long et~al.(2015)Long, Shelhamer and Darrell}]{long2015fully}
\bibinfo{author}{Long, J.}, \bibinfo{author}{Shelhamer, E.},
  \bibinfo{author}{Darrell, T.}, \bibinfo{year}{2015}.
\newblock \bibinfo{title}{Fully convolutional networks for semantic
  segmentation}, in: \bibinfo{booktitle}{IEEE CVPR Proceedings}, pp.
  \bibinfo{pages}{3431--3440}.
\bibitem[{Lucas and Kanade(1981)}]{Lucas1981}
\bibinfo{author}{Lucas, B.D.}, \bibinfo{author}{Kanade, T.},
  \bibinfo{year}{1981}.
\newblock \bibinfo{title}{{An Iterative Image Registration Technique with an
  Application to Stereo Vision}}.
\newblock \bibinfo{journal}{Imaging} \bibinfo{volume}{130},
  \bibinfo{pages}{674--679}.
\newblock \DOIprefix\doi{10.1109/HPDC.2004.1323531}.
\bibitem[{Mansilla and Ferrante(2018)}]{mansilla2018}
\bibinfo{author}{Mansilla, L.}, \bibinfo{author}{Ferrante, E.},
  \bibinfo{year}{2018}.
\newblock \bibinfo{title}{Segmentaci{\'o}n multi-atlas de im{\'a}genes
  m{\'e}dicas con selecci{\'o}n de atlas inteligente y control de calidad
  autom{\'a}tico}, in: \bibinfo{booktitle}{XXIV Congreso Argentino de Ciencias
  de la Computaci{\'o}n (La Plata, 2018).}, pp. \bibinfo{pages}{371--380}.
\bibitem[{Marstal et~al.(2016)Marstal, Berendsen, Staring and
  Klein}]{simpleelastix}
\bibinfo{author}{Marstal, K.}, \bibinfo{author}{Berendsen, F.},
  \bibinfo{author}{Staring, M.}, \bibinfo{author}{Klein, S.},
  \bibinfo{year}{2016}.
\newblock \bibinfo{title}{Simpleelastix: A user-friendly, multi-lingual library
  for medical image registration}, in: \bibinfo{booktitle}{Proceedings of the
  IEEE CVPR Workshops}, pp. \bibinfo{pages}{134--142}.
\bibitem[{Oh and Kim(2017)}]{oh2017deformable}
\bibinfo{author}{Oh, S.}, \bibinfo{author}{Kim, S.}, \bibinfo{year}{2017}.
\newblock \bibinfo{title}{Deformable image registration in radiation therapy}.
\newblock \bibinfo{journal}{Radiation oncology journal} \bibinfo{volume}{35},
  \bibinfo{pages}{101}.
\bibitem[{Oktay et~al.(2018)Oktay, Ferrante et~al.}]{Oktay2017}
\bibinfo{author}{Oktay, O.}, \bibinfo{author}{Ferrante, E.}, et~al.,
  \bibinfo{year}{2018}.
\newblock \bibinfo{title}{Anatomically constrained neural networks (acnns):
  application to cardiac image enhancement and segmentation}.
\newblock \bibinfo{journal}{IEEE TMI} \bibinfo{volume}{37},
  \bibinfo{pages}{384--395}.
\bibitem[{Paragios et~al.(2016)}]{Paragios2016}
\bibinfo{author}{Paragios, N.}, et~al., \bibinfo{year}{2016}.
\newblock \bibinfo{title}{(hyper)-graphical models in biomedical image
  analysis}.
\newblock \bibinfo{journal}{Medical Image Analysis} \bibinfo{volume}{33},
  \bibinfo{pages}{102 -- 106}.
\bibitem[{Ren et~al.(2017)Ren, Yan, Ni, Liu, Yang and
  Zha}]{ren2017unsupervised}
\bibinfo{author}{Ren, Z.}, \bibinfo{author}{Yan, J.}, \bibinfo{author}{Ni, B.},
  \bibinfo{author}{Liu, B.}, \bibinfo{author}{Yang, X.}, \bibinfo{author}{Zha,
  H.}, \bibinfo{year}{2017}.
\newblock \bibinfo{title}{Unsupervised deep learning for optical flow
  estimation.}, in: \bibinfo{booktitle}{AAAI}, p.~\bibinfo{pages}{7}.
\bibitem[{Roh{\'e} et~al.(2017)Roh{\'e}, Datar, Heimann, Sermesant and
  Pennec}]{Rohe2017}
\bibinfo{author}{Roh{\'e}, M.M.}, \bibinfo{author}{Datar, M.},
  \bibinfo{author}{Heimann, T.}, \bibinfo{author}{Sermesant, M.},
  \bibinfo{author}{Pennec, X.}, \bibinfo{year}{2017}.
\newblock \bibinfo{title}{Svf-net: Learning deformable image registration using
  shape matching}, in: \bibinfo{booktitle}{MICCAI},
  \bibinfo{organization}{Springer}. pp. \bibinfo{pages}{266--274}.
\bibitem[{Ronneberger et~al.(2015)Ronneberger, Fischer and
  Brox}]{Ronneberger2015}
\bibinfo{author}{Ronneberger, O.}, \bibinfo{author}{Fischer, P.},
  \bibinfo{author}{Brox, T.}, \bibinfo{year}{2015}.
\newblock \bibinfo{title}{U-net: Convolutional networks for biomedical image
  segmentation}, in: \bibinfo{booktitle}{MICCAI},
  \bibinfo{organization}{Springer}. pp. \bibinfo{pages}{234--241}.
\bibitem[{Shakeri et~al.(2016)Shakeri, Ferrante, Tsogkas, Lippe, Kadoury,
  Kokkinos and Paragios}]{shakeri2016prior}
\bibinfo{author}{Shakeri, M.}, \bibinfo{author}{Ferrante, E.},
  \bibinfo{author}{Tsogkas, S.}, \bibinfo{author}{Lippe, S.},
  \bibinfo{author}{Kadoury, S.}, \bibinfo{author}{Kokkinos, I.},
  \bibinfo{author}{Paragios, N.}, \bibinfo{year}{2016}.
\newblock \bibinfo{title}{Prior-based coregistration and cosegmentation}, in:
  \bibinfo{booktitle}{MICCAI}, \bibinfo{organization}{Springer}. pp.
  \bibinfo{pages}{529--537}.
\bibitem[{Shiraishi et~al.(2000)Shiraishi, Katsuragawa, Ikezoe, Matsumoto,
  Kobayashi, Komatsu, Matsui, Fujita, Kodera and Doi}]{JSRT}
\bibinfo{author}{Shiraishi, J.}, \bibinfo{author}{Katsuragawa, S.},
  \bibinfo{author}{Ikezoe, J.}, \bibinfo{author}{Matsumoto, T.},
  \bibinfo{author}{Kobayashi, T.}, \bibinfo{author}{Komatsu, K.i.},
  \bibinfo{author}{Matsui, M.}, \bibinfo{author}{Fujita, H.},
  \bibinfo{author}{Kodera, Y.}, \bibinfo{author}{Doi, K.},
  \bibinfo{year}{2000}.
\newblock \bibinfo{title}{Development of a digital image database for chest
  radiographs with and without a lung nodule: receiver operating characteristic
  analysis of radiologists' detection of pulmonary nodules}.
\newblock \bibinfo{journal}{American Journal of Roentgenology}
  \bibinfo{volume}{174}, \bibinfo{pages}{71--74}.
\bibitem[{Sokooti et~al.(2017)}]{Sokooti2017}
\bibinfo{author}{Sokooti, H.}, et~al., \bibinfo{year}{2017}.
\newblock \bibinfo{title}{Nonrigid image registration using multi-scale 3d
  convolutional neural networks}, in: \bibinfo{booktitle}{MICCAI},
  \bibinfo{organization}{Springer}. pp. \bibinfo{pages}{232--239}.
\bibitem[{Sotiras et~al.(2013)Sotiras, Davatzikos and Paragios}]{Sotiras2013}
\bibinfo{author}{Sotiras, A.}, \bibinfo{author}{Davatzikos, C.},
  \bibinfo{author}{Paragios, N.}, \bibinfo{year}{2013}.
\newblock \bibinfo{title}{Deformable medical image registration: A survey}.
\newblock \bibinfo{journal}{IEEE TMI} \bibinfo{volume}{32},
  \bibinfo{pages}{1153--1190}.
\bibitem[{Stergios et~al.(2018)Stergios, Mihir, Maria, Guillaume, Marie-Pierre,
  Stavroula and Nikos}]{christodoulidis2018}
\bibinfo{author}{Stergios, C.}, \bibinfo{author}{Mihir, S.},
  \bibinfo{author}{Maria, V.}, \bibinfo{author}{Guillaume, C.},
  \bibinfo{author}{Marie-Pierre, R.}, \bibinfo{author}{Stavroula, M.},
  \bibinfo{author}{Nikos, P.}, \bibinfo{year}{2018}.
\newblock \bibinfo{title}{Linear and deformable image registration with 3d
  convolutional neural networks}, in: \bibinfo{booktitle}{Image Analysis for
  Moving Organ, Breast, and Thoracic Images}. \bibinfo{publisher}{Springer},
  pp. \bibinfo{pages}{13--22}.
\bibitem[{Valindria et~al.(2017)Valindria, Lavdas, Bai, Kamnitsas, Aboagye,
  Rockall, Rueckert and Glocker}]{valindria2017reverse}
\bibinfo{author}{Valindria, V.V.}, \bibinfo{author}{Lavdas, I.},
  \bibinfo{author}{Bai, W.}, \bibinfo{author}{Kamnitsas, K.},
  \bibinfo{author}{Aboagye, E.O.}, \bibinfo{author}{Rockall, A.G.},
  \bibinfo{author}{Rueckert, D.}, \bibinfo{author}{Glocker, B.},
  \bibinfo{year}{2017}.
\newblock \bibinfo{title}{Reverse classification accuracy: predicting
  segmentation performance in the absence of ground truth}.
\newblock \bibinfo{journal}{IEEE TMI} \bibinfo{volume}{36},
  \bibinfo{pages}{1597--1606}.
\bibitem[{Viergever et~al.(2016)Viergever, Maintz, Klein, Murphy, Staring and
  Pluim}]{Vierger2016}
\bibinfo{author}{Viergever, M.A.}, \bibinfo{author}{Maintz, J.A.},
  \bibinfo{author}{Klein, S.}, \bibinfo{author}{Murphy, K.},
  \bibinfo{author}{Staring, M.}, \bibinfo{author}{Pluim, J.P.},
  \bibinfo{year}{2016}.
\newblock \bibinfo{title}{A survey of medical image registration – under
  review}.
\newblock \bibinfo{journal}{Medical Image Analysis} \bibinfo{volume}{33},
  \bibinfo{pages}{140--144}.
\newblock \URLprefix
  \url{http://www.sciencedirect.com/science/article/pii/S1361841516301074},
  \DOIprefix\doi{https://doi.org/10.1016/j.media.2016.06.030}.
  \bibinfo{note}{20th anniversary of the Medical Image Analysis journal
  (MedIA)}.
\bibitem[{Vincent et~al.(2010)Vincent, Larochelle, Lajoie, Bengio and
  Manzagol}]{vincent2010stacked}
\bibinfo{author}{Vincent, P.}, \bibinfo{author}{Larochelle, H.},
  \bibinfo{author}{Lajoie, I.}, \bibinfo{author}{Bengio, Y.},
  \bibinfo{author}{Manzagol, P.A.}, \bibinfo{year}{2010}.
\newblock \bibinfo{title}{Stacked denoising autoencoders: Learning useful
  representations in a deep network with a local denoising criterion}.
\newblock \bibinfo{journal}{JMLR} \bibinfo{volume}{11},
  \bibinfo{pages}{3371--3408}.
\bibitem[{de~Vos et~al.(2019)de~Vos, Berendsen, Viergever, Sokooti, Staring and
  I{\v{s}}gum}]{de2019deep}
\bibinfo{author}{de~Vos, B.D.}, \bibinfo{author}{Berendsen, F.F.},
  \bibinfo{author}{Viergever, M.A.}, \bibinfo{author}{Sokooti, H.},
  \bibinfo{author}{Staring, M.}, \bibinfo{author}{I{\v{s}}gum, I.},
  \bibinfo{year}{2019}.
\newblock \bibinfo{title}{A deep learning framework for unsupervised affine and
  deformable image registration}.
\newblock \bibinfo{journal}{Medical Image Analysis} \bibinfo{volume}{52},
  \bibinfo{pages}{128--143}.
\bibitem[{de~Vos et~al.(2017)de~Vos, Berendsen, Viergever, Staring and
  I{\v{s}}gum}]{DeVos2017}
\bibinfo{author}{de~Vos, B.D.}, \bibinfo{author}{Berendsen, F.F.},
  \bibinfo{author}{Viergever, M.A.}, \bibinfo{author}{Staring, M.},
  \bibinfo{author}{I{\v{s}}gum, I.}, \bibinfo{year}{2017}.
\newblock \bibinfo{title}{End-to-end unsupervised deformable image registration
  with a convolutional neural network}, in: \bibinfo{booktitle}{DLMIA, ML-CDS}.
  \bibinfo{publisher}{Springer}, pp. \bibinfo{pages}{204--212}.
\bibitem[{Wang et~al.(2017)Wang, Peng, Lu, Lu, Bagheri and
  Summers}]{wang2017chestx}
\bibinfo{author}{Wang, X.}, \bibinfo{author}{Peng, Y.}, \bibinfo{author}{Lu,
  L.}, \bibinfo{author}{Lu, Z.}, \bibinfo{author}{Bagheri, M.},
  \bibinfo{author}{Summers, R.M.}, \bibinfo{year}{2017}.
\newblock \bibinfo{title}{Chestx-ray8: Hospital-scale chest x-ray database and
  benchmarks on weakly-supervised classification and localization of common
  thorax diseases}, in: \bibinfo{booktitle}{IEEE CVPR Proceedings}, pp.
  \bibinfo{pages}{2097--2106}.
\bibitem[{Wouters et~al.(2006)Wouters, D'Agostino, Maes, Vandermeulen and
  Suetens}]{Wouters2006}
\bibinfo{author}{Wouters, J.}, \bibinfo{author}{D'Agostino, E.},
  \bibinfo{author}{Maes, F.}, \bibinfo{author}{Vandermeulen, D.},
  \bibinfo{author}{Suetens, P.}, \bibinfo{year}{2006}.
\newblock \bibinfo{title}{Non-rigid brain image registration using a
  statistical deformation model}, in: \bibinfo{booktitle}{Medical Imaging 2006:
  Image Processing}, \bibinfo{organization}{International Society for Optics
  and Photonics}. p. \bibinfo{pages}{614411}.
\bibitem[{Yang et~al.(2017)}]{Yang2017}
\bibinfo{author}{Yang, X.}, et~al., \bibinfo{year}{2017}.
\newblock \bibinfo{title}{Quicksilver: Fast predictive image registration--a
  deep learning approach}.
\newblock \bibinfo{journal}{NeuroImage} \bibinfo{volume}{158},
  \bibinfo{pages}{378--396}.

\end{thebibliography}

\end{document}